\documentclass[12pt]{article}
\usepackage{
amsmath,latexsym}

\newcommand{\sect}[1]{\setcounter{equation}{0}\section{#1}}

\newcommand{\eq}{\begin{equation}}

\newcommand{\en}{\end{equation}}

\newcommand{\enn}{\nonumber \end{equation}}

\def\epsihat{{\widehat{\varepsilon}}}
\def\epsilonhat{{\widehat{\epsilon}}}
\def\deltahat{ {\widehat\delta} }
\def\Omhat{\widehat{\Om}}
\def\Vhat{\widehat{V}}

\def\Vtildehat{{\widehat{\Vtilde}}}
\def\Rtildehat{{\widehat{\Rtilde}}}

\def\epsitildehat{\widehat{\epsitilde}}

\def\sk{\vskip .4cm}
\def\noi{\noindent}
\def\om{\omega}

\def\ga{\gamma}

\let \part\partial

\def\unquarto{{1 \over 4}}

\def\unmezzo{{1 \over 2}}
\def\epsi{\varepsilon}
\def\we{\wedge}

\def\de{\delta}

\def\part{\partial}

\def\sk{\vskip .4cm}

\def\noi{\noindent}

\def\X0{X^0}

\def\om{\omega}

\def\ga{\gamma}

\def\unquarto{{1 \over 4}}
\def\unmezzo{{1 \over 2}}
\def\epsi{\varepsilon}

\def\we{\wedge}

\def\de{\delta}

\def\Rhat#1#2{ \Rh^{#1}_{~~~#2} }

\def\LL#1#2#3#4#5#6#7#8{\La^{~#2~#4}_{#1~#3}|^{#5~#7}_{~#6~#8}}

\def\square{{\,\lower0.9pt\vbox{\hrule \hbox{\vrule height 0.2 cm
\hskip 0.2 cm \vrule height 0.2 cm}\hrule}\,}}

\def\westar{\we_\star}

\def\omtilde{\tilde \om}
\def\Vtilde{\widetilde{V}}

\def\Rtilde{\widetilde{R}}

\def\epsitilde{\widetilde{\epsi}}

\def\Om{\Omega}

\def\Rhat{\widehat{R}}
\def\omhat{\widehat{\om}}
\def\omtildehat{\widehat{\omtilde}}

\def\Phihat{\widehat{\Phi}}

\def\nn{\nonumber}

\def\LL{I\!\!L}

\def\sA{^{}_{[A}}
\def\Bs{^{}_{B]}}

\def\As]{^{}_{A]}}

\begin{document}

\begin{titlepage}
\begin{center}
{\Large \bf Noncommutative  gravity at second order\\[.4em] 
via Seiberg-Witten  map}
\\[3em]
{\large {\bf Paolo Aschieri$^1$}, {\bf Leonardo Castellani$^1$} and  {\bf Marija Dimitrijevi\'c$^2$}} \\ [2em] {\sl $^1$Dipartimento di Scienze e Innovazione Tecnologica
\\ INFN Gruppo collegato di Alessandria,\\Universit\`a del Piemonte Orientale,\\ Viale T. Michel 11,  15121 Alessandria, Italy\\[1em]
$^2$University of Belgrade, Faculty of Physics\\ Studentski trg 12, 11000
Beograd, Serbia
}\\ [4em]
\end{center}

\begin{abstract}

\vskip 0.2cm
We develop a general strategy to express noncommutative
actions in terms of commutative ones by using a recently developed 
geometric generalization of the  Seiberg-Witten  map (SW map) between
noncommutative and commutative fields.  

We apply this general scheme to the noncommutative vierbein gravity action 
and provide a SW differential equation for the action itself as well
as a recursive solution at all orders in the noncommutativity
parameter $\theta$. We thus express the action at order
$\theta^{n+2}$ in terms of noncommutative fields of order at most
$\theta^{n+1}$ and, iterating the procedure, in terms of
noncommutative fields of order at most $\theta^n$.

This in particular provides the explicit expression of the 
action at order
$\theta^2$ in terms of the usual commutative spin
connection and  vierbein fields. 
The result is an extended gravity action on commutative
spacetime that is manifestly invariant under local Lorentz rotations
and general coordinate transformations.

 \end{abstract}

\vskip 5cm \noi \hrule \vskip.2cm \noi {\small aschieri@to.infn.it\\
leonardo.castellani@mfn.unipmn.it\\ dmarija@ipb.ac.rs}

\end{titlepage}

\newpage
\setcounter{page}{1}

\sect{Introduction}
Generalizations and extensions of Einstein gravity have a long history, with motivations that are both
theoretical and experimental, see for example the reviews in \cite{EGTreview}. In the last decade 
numerous higher order extensions of the Einstein action have been considered, mostly
for applications in cosmological phenomenology, related to the issues
of dark matter and dark energy.  

A recent way to obtain higher order gravity theories is based on
noncommutativity of spacetime. 
Gravity on noncommutative spacetime may be expected to capture some aspects of quantum
gravity since there are indications that at short distances spacetime indeed
becomes noncommutative (see for example \cite{example} and references therein).
Thus an extended higher order gravity theory on commutative spacetime, 
obtained from gravity on noncommutative spacetime, could be seen as an
effective theory of a more fundamental quantum theory.

We  present a manifestly Lorentz invariant expression for the first nontrivial
term of the extended gravity theory obtained from the  noncommutative vierbein gravity studied in
  \cite{AC1, AC3}.  The resulting action is geometric and hence invariant
  under general  coordinate transformations. It predicts 
specific higher curvature couplings, whose theoretical and phenomenological
properties are ripe for being investigated and compared with other
extended gravity approaches.
\sk
\sk
\noi{\it $\star$-Product noncommutativity}\\[.4em]\noi
Commutative spacetime can be described via the commutative
algebra of complex valued functions on spacetime.
Noncommutative spacetime can be described by a noncommutative deformation
of this algebra.
One way of deforming the commutative product of functions is 
by means of a  $\star$-product. 
We consider $\star$-products originating from twist deformation: 
in this case the deformation depends on a dimensionful parameter $\theta$ that
is a constant  antisymmetric matrix with components $\theta^{AB}$, and
on a set of commuting vector fields $X^A$. 
It generalizes the Moyal-Groenewold \cite{MoyalGroenewold} $\star$-product 
between phase-space functions. The $\star$-product
between functions (i.e. $0$-forms), and in general between  arbitrary
exterior forms (the $\star$-deformed exterior product), is defined 
by :
   \begin{flalign}
    \tau \westar \tau' &\equiv  \sum_{n=0}^\infty \left({i \over 2}\right)^n \theta^{A_1B_1} \cdots \theta^{A_nB_n}
   (\ell_{X_{A_1}} \cdots \ell_{X_{A_n}} \tau) \we  (\ell_{X_{B_1}} \cdots \ell_{X_{B_n}} \tau')  \nonumber \\
  &= \tau \we \tau' + {i \over 2} \theta^{AB} (\ell_{X_A} \tau) \we (\ell_{X_B} \tau') + {1 \over 2!}  {\left( i \over 2 \right)^2} \theta^{A_1B_1} \theta^{A_2B_2}  (\ell_{X_{A_1}} \ell_{X_{A_2}} \tau) \we
 (\ell_{X_{B_1}} \ell_{X_{B_2}} \tau') + \cdots  
  \label{defwestar}
  \end{flalign}
       \noi where $\ell_{X_A}$ are Lie derivatives along commuting
       vector fields $X_A$. This product is noncommutative in
       the regions of spacetime where the vector fields $X_A$ are
       nonvanishing, and is associative due to $[X_A , X_B]=0$. 

If  spacetime is flat Minkowski space and the vector
fields $X_A$ are chosen to  coincide with the partial derivatives
$\partial_\mu$, and if $\tau$, $\tau'$ are $0$-forms, then $\tau\star\tau'$ reduces to the well-known Moyal-Groenewold product
\cite{MoyalGroenewold}. In particular for  coordinate functions we
have $[x^\mu , x^\nu]_\star\equiv x^\mu\star x^\nu- x^\nu\star
x^\mu=i\theta^{\mu\nu} $ (i.e. we have constant noncommutativity,
since $\theta$ is constant).
The more general $\star$-product (\ref{defwestar}) is given by a
choice of spacetime dependent and commuting vector fields
$X_A=X_A^\mu(x)\partial_\mu$ 
that may be vanishing outside a given region (we refer to
\cite{NCG} Sec. 2.3  and \cite{Waldmann}
  Sec. 5.2 for examples).
In this case noncommutativity of the coordinates is spacetime
dependent: $[x^\mu , x^\nu]_\star=i\theta^{AB}X_A^\mu(x)X_B^\nu(x)\equiv i\Theta^{\mu\nu}(x)$.
Further properties of  $\star$-products originating from abelian twists
${\cal F}=e^{-\frac{i}{2}\theta^{AB}X_A\otimes X_B},$ and of the
associated twist differential geometry are summarized  for example in
Appendix A of \cite{AC3}.
\sk
\sk
\noi{\it $\star$-Gauge theory and Seiberg-Witten map (SW map)}\\[.4em]\noi
Noncommutative gauge theories are obtained from
nonabelian ones on commutative space by substituting ordinary
products between fields and forms with $\star$-products.
This recipe  yields NC actions invariant under deformations of the
original gauge symmetry, but brings into play many more fields than those present in the original action.
This is easily understood: consider for example the $\star$-Yang-Mills field strength:
 \eq
F_{\mu\nu}^{I}T_{I} = \part_{\mu} A_{\nu}^{I}T_{I}-\part_{\nu} A_{\mu}^{I}T_{I}-(A_{\mu}^{J} \star A_{\nu}^{K}-
A_{\nu}^{J} \star A_{\mu}^{K})T_{J}T_{K}~.
\en
Because of noncommutativity of the $\star$-product, anticommutators as well as commutators of
group generators are appearing in the right-hand side. 
Since $T_JT_K$ must be a linear combination of $T_{I}$'s, we see
we cannot in general consider the generators $T_I$ to be a basis of the Lie
algebra of the gauge group $G$. We have to enlarge the original set of
Lie algebra generators and  to consider the generators $T_I$ to be a basis
for the whole universal enveloping algebra of $G$. The range of the index $I$ increases,
and so does the number of independent field components
$A^{I}_{\mu}$. One can reduce this proliferation by choosing a
specific representation for the generators $T_{I}$ (for 
$SU(2)$ we can take its generators to be
the Pauli matrices, so that a basis for the enveloping algebra only requires an additional matrix proportional to the unit matrix).

Even the remaining extra degrees of freedom can be eliminated, by use
of the Seiberg-Witten map (SW map), 
relating the enveloping algebra valued gauge potential $A_\mu^IT_I$ to the original Lie
algebra valued potential of the undeformed theory, the so-called
classical gauge fields. The SW map applies also to matter fields, so
that for example noncommutative matter fields in the adjoint $\Phi^IT_I$
are expressed in terms of the fewer Lie algebra valued commutative ones. 

\sk
The  SW map \cite{SW} was initially developed for $U(N)$ gauge fields
and Moyal-Groenewold $\star$-products (constant noncommutativity), then
applied
to gauge fields of arbitrary gauge groups and extended to matter fields
in \cite{Madore,  Jurco1, Jurco2}.
An explicit solution for $U(1)$ gauge
theory was presented in \cite{Ooguri} developing earlier work
\cite{Mehen:2000vs, Liu:2000mja, JSW}. 
Concerning  nonabelian gauge
groups an iterative procedure, based on recurrence relations, was devised in
 \cite{Ulker} (improving on results of \cite{Bichl:2001cq}, Sec. 6);
 it allows to construct  the SW map as a power series expansion in
 $\theta$ in the particular case of Moyal-Groenewold $\star$-product.
 
The SW map  is also mathematically very rich and was obtained 
as a power series in $\theta$ expansion for $U(1)$ gauge fields in \cite{JSW}  in case of an arbitrary
$\star$-product  (originating from an arbitrary Poisson tensor,
i.e., nonconstant noncommutativity) using
Kontsevich's results \cite{Kontsevich:1997vb};  in the nonabelian case the
situation (for nonconstant noncommutativity) is more 
involved and there is no definite result (despite
 interesting partial ones \cite{JS}). 

In \cite{AC3} we gave a geometric formulation of SW map and
generalized the recurrence relations presented in  \cite{Ulker} to the case of  
$\star$-products obtained via a set of mutually commuting vector
fields.
These vector fields can be spacetime dependent and hence
we obtained a SW map for nonabelian gauge fields with nonconstant
noncommutativity on arbitrary (spacetime) manifolds. 
\sk

It is now conceptually easy to obtain commutative actions from
noncommutative ones: {\it i$_{\,}$}) Consider the noncommutative gauge and matter fields as
dependent on the commutative ones via the SW map; {\it ii$_{\,}$}) Expand the star
products and the  NC fields in power series of the noncommutativity
parameter $\theta$. 

The result is an action that contains higher order corrections in the field
strenght, its derivatives and derivatives of the matter fields, organized in a power series in $\theta$. 
Every power in $\theta$ is separately invariant under ordinary
gauge transformations, since the NC action is invariant under NC gauge
transformations, and these latter are induced, via the SW map, by
ordinary  gauge transformations  (that do not depend on $\theta$) on
the classical fields.

\sk
We have applied this strategy to vierbein gravity (where the gauge group is the local Lorentz group) coupled to fermions
\cite{AC3} (reviewed in \cite{NCGF}), to gauge fields \cite{AC4} and to scalars \cite{AC5}. 
In ref. \cite{AC3}  the second order correction (in $\theta$) to pure
vierbein gravity was presented in terms of first order fields. 
Here  we give it in terms of classical fields and in a manifestly
Lorentz gauge invariant form. 
\sk
\sk
\noi{\it Plan of the paper}\\[.4em]\noi
In Section 2 we recall the geometric action for pure NC vierbein
gravity. Section 3 deals with the geometric SW map for
$\star$-products  with commuting vector fields ($\star$-products from
abelian twists).  In this section we consider an arbitrary
gauge theory with nonabelian gauge group.  We recall the SW
differential equations and the associated recursive relations
expressing the fields
at order $\theta^{n+1}$ in terms of fields 
of order at most $\theta^n$.  Techniques for the calculation of the  SW
differential equation and recursive relations for composite fields
are then provided.

In Section 4 we apply these results to the noncommutative vierbein
gravity action. We establish the SW differential equation it satisfies
and express the action at order $\theta^{n+2}$ in terms of
the noncommutative spin connection and matter fields up to order
$\theta^{n+1}$, and, iterating the procedure, in terms of  the fields
up to order $\theta^n$. In Section 5 we present the $\theta^2$  correction of NC
vierbein gravity, and of the NC cosmological term, in terms of classical fields.

In the appendices we list the Cartan
formulae used throughout the main text and we summarize the $D=4$ gamma matrix conventions.

\sect{NC vierbein gravity action}

The noncommutative action reads \cite{AC1}:
\eq
  S_{NC} =  \int Tr \big(i \ga_5 \Rhat \westar \Vhat \westar \Vhat  \big)~,
 \label{NCaction}
\en
\noi where  the  curvature $\Rhat (\Omhat)$ is defined in terms of the NC spin connection as
 \eq
\Rhat = d \Omhat- \Omhat \westar \Omhat~.
 \en
This definition implies the Bianchi identity:
 \eq
 D \Rhat \equiv  d \Rhat - \Omhat \westar \Rhat +  \Rhat  \westar \Omhat =0~.
 \en

\noi The NC vierbein $\Vhat$, spin connection $\Omhat$ and curvature
$\Rhat$ are valued  in Dirac gamma matrices and the trace in the above
action is taken on their spinor indices. 
The Dirac gamma matrices expansion of the NC fields is \cite{Chamseddine}:
 \begin{flalign}
 &  \Omhat = {1 \over 4} \omhat^{ab}  \ga_{ab}  + i \omhat 1 + \omtildehat \ga_5~,  \label{Omhat}\\
 &  \Vhat =  \Vhat^{a} \ga_a  + \Vtildehat{}^a \ga_a \ga_5~, \\
& 
 \Rhat= {1 \over 4} \Rhat^{ab}  \ga_{ab}  +i \Rhat  +
\Rtildehat \ga_5 ~. 
\end{flalign}
 The classical limits of the NC fields are
constrained to be the classical fields \cite{AC1,MiaoZhang, AC3}:
 \eq
  \Om \equiv \unquarto \om^{ab} \ga_{ab}~, ~~~V \equiv V^a \ga_a~,~~~R \equiv \unquarto R^{ab} \ga_{ab}~ \label{classicalfields}
 \en
\noi with
 \eq
 R^{ab} = d \om^{ab} - \om^{ac} \we \om^{cb}~
 \en
where repeated Lorentz indices are summed with the Minkowski metric $\eta_{ab}$.
\noi By recalling that $  Tr (\ga_{ab} \ga_c \ga_d \ga_5 ) = -4 i \epsi_{abcd}$ one can easily check that the classical limit of the NC action (\ref{NCaction}) (obtained by replacing
NC fields by classical fields, and deformed exterior products by ordinary
exterior products) reproduces the usual first order formulation of the
Einstein-Hilbert action:
 \eq
 S_{commutative} =   \int Tr \big(i \ga_5 R \wedge V \wedge V  \big)=\int R^{ab} \we V^c \we V^d \epsi_{abcd}~.
 \en

\subsection{Noncommutative symmetries}

 The NC action (\ref{NCaction}) is invariant under general coordinate transformations (being the integral of a 
 4-form) and under the $\star$-gauge variations:
\begin{flalign}\
 {\deltahat_\epsihat}
 \Omhat&=d \epsihat - \Omhat \star \epsihat + \epsihat \star \Omhat ~~\Longrightarrow~~
{\deltahat_\epsihat}
 \Rhat = -\Rhat  \star \epsihat + \epsihat \star
 \Rhat ~,\nonumber \\
  \deltahat_\epsihat \Vhat &= -\Vhat \star \epsihat + \epsihat \star
  \Vhat \label{stargauge} 
   \end{flalign}
   with an arbitrary parameter $\epsihat_{}(x)$ commuting with $\ga_5$,
   i.e., valued in the even gamma matrix algebra,
\eq
\epsilonhat =  {1 \over 4} \epsihat^{ab}  \ga_{ab}  +  i \epsihat + \epsitildehat \ga_5 ~.
\en
The invariance of the noncommutative action under these transformations
relies on the cyclicity of the integral (and of the trace) and on
 $\widehat\epsi$ commuting with $\ga_5$. The classical limit of the gauge
 parameter is constrained to be:
 \eq
 \epsi = \unquarto \epsi^{ab} \ga_{ab}
 \en
 \noi and one can check that the classical limit of the gauge
 transformations (\ref{stargauge}) reproduces the usual local Lorentz
 rotations on the vielbein and on the spin connection.

As discussed in the introduction, the extra fields entering in
the expansions of $\Omhat$, $\Vhat$ and $\epsihat$ are due to
the noncommutativity of the star product.
Indeed the $\star$-gauge variations of the fields 
(\ref{stargauge}) include also anticommutators of gamma
matrices. Since for example the anticommutator
 $\{ \ga_{ab},\ga_{cd} \}$ contains $1$ and $\ga_5$, we see that the corresponding fields
 must be included in the expansion of $\Omhat$. Similarly, $\Vhat$ must contain a $\ga_a \ga_5$ term due
 to $\{ \ga_{ab},\ga_{c} \}$.

\sk
All the components along the $SO(1,3)$ enveloping algebra generators are taken to be real, and therefore fields and curvatures satisfy the hermiticity properties: 
 \eq
 \Omhat^\dagger = - \ga_0 \Omhat \ga_0~,~~~\Vhat^\dagger = \ga_0 \Vhat \ga_0~,~~~\Rhat^\dagger =- \ga_0 \Rhat \ga_0~,
 \en 
\noi i.e., $\Omhat$ and $\Rhat$ are $\ga_0$-antihermitian, while $\Vhat$ is $\ga_0$-hermitian. Using these rules it is a quick matter to check that the noncommutative action
(\ref{NCaction}) is real. 

\sect{Geometric Seiberg-Witten map }
The Seiberg-Witten map relates the noncommutative gauge field $\Omhat$
to the ordinary  $\Om$, and the noncommutative gauge parameter
$\epsihat$ to  the ordinary  $\epsi$  so as to 
satisfy:
 \eq
\Omhat (\Om) + {\widehat{\delta}}_\epsihat
\Omhat (\Om) = \Omhat (\Om + \de_\epsi \Om)    
\label{SWcondition}
\en
 with 
  \begin{flalign} 
   & 
  \de_\epsi \Om_\mu = \part_\mu \epsi + \epsi \Om_\mu -  \Om_\mu 
      \epsi~, \\
      & 
  \deltahat_\epsihat{} \Omhat_\mu = \part_\mu \epsihat + \epsihat \star \Omhat_\mu -  \Omhat_\mu \star     
      \epsihat~.
      \end{flalign}
   
\noi  In words: the dependence of the noncommutative gauge field on the ordinary one is fixed
by requiring that ordinary gauge variations of $\Om$ inside $\Omhat(\Om)$ produce the noncommutative
gauge variation of $\Omhat$.

Similarly noncommutative matter fields are related to the commutative
ones by requiring
 \eq
\Phihat(\Phi,\Om) + {\widehat{\delta}}_\epsihat
\Phihat (\Phi,\Om) = \Phihat (\Phi + \de_\epsi \Phi, \Om+\delta_\epsi\Omega) ~.   
\label{SWcondition1}
\en

The conditions (\ref{SWcondition}), (\ref{SWcondition1}) are implied by the following differential equations in the
noncommutativity parameter $\theta^{AB}$ \cite{SW, AC3}:
\begin{flalign}
 &  { \part ~~\over \part \theta^{AB}} \Omhat = {i \over 4}  \{\Omhat^{}_{[A},  \ell^{}_{B]} \Omhat +\Rhat^{}_{B]}\}_\star ~,\label{diffeqOm} \\
&   { \part ~~\over \part \theta^{AB}}  \Phihat = {i \over 4}  \{\Omhat_{[A},  \LL^{}_{B]} \Phihat \}_\star ~,\label{diffeqphi} \\
&   { \part ~~\over \part \theta^{AB}}  \epsihat = {i \over 4}  \{\Omhat_{[A},  \ell^{}_{B]} \epsihat \}_\star ~,\label{diffeqepsi} 
\end{flalign}
where:\\
$\bullet$ The bracket $[A\, B]$ denotes that the indices $A$ and $B$ are
antisymmetrized with weight 1, so that for example 
$\Omhat\sA \Rhat\Bs=\frac{1}{2}(\Omhat_A\Rhat_B-\Omhat_B\Rhat_A)$.
 The bracket $\{~,~\}_\star$ is the usual $\star$-anticommutator, for
example
$\{\Om_A,R_B\}_\star=\Om_A\star R_B+R_B\star\Om_A$. \\[.2em]
$\bullet$ $\ell_B$ is the Lie derivative along the 
vector field $X_B$.\\[.2em]
$\bullet$ $\Omhat_A$, $\Rhat_A$ are defined as the contraction $i_A$ along the tangent
vector $X_A$ of the exterior forms $\Omhat$, $\Rhat$, i.e. $\Omhat_A\equiv i_A\Omhat$,
$\Rhat_A \equiv i_A \Rhat$.\\[.2em]
$\bullet$ The second differential equation holds for  fields transforming
in the adjoint representation. Notice that $\Phihat$ can  also be an
exterior form. The notation $\LL_B$ is defined by 
$\,\LL_B\equiv \ell_B+L_B\,$ where $L_B$ is the  covariant Lie derivative  along the
tangent vector $X_B$; it acts on the field $\Phihat$ as 
$$L_B  \Phihat = \ell_B \Phihat - [\Omhat_B , \Phihat]_\star\;,$$ where  
$[\Omhat_B , \Phihat]_\star=\Omhat_B \star\Phihat- \Phihat\star\Omhat_B$.
In fact the covariant Lie derivative $L_B$ can be written in Cartan form:
 \eq
  L_B = i_B D + D i_B~,
    \en
where $D$ is the covariant derivative.
\sk
The differential equations (\ref{diffeqOm})-(\ref{diffeqepsi}) hold for any abelian twist defined by arbitrary commuting vector
fields $X_A$  (that can vanish in some region of
spacetime)  \cite{AC3}.  They reduce to the usual Seiberg-Witten
differential equations formulae \cite{SW}  for an
arbitrary nonabelian gauge group in case the
twist becomes a Moyal-Groenewold  twist, i.e., $X_A\to \partial_\mu$.

\sk

We solve these differential equations order by order in $\theta$
by expanding $\Omhat$ and $\Phihat$ in power series of $\theta$
\eq
\Omhat=\Om^0 + \Om^1+\Om^2\;\ldots+\Om^n\ldots~~,~~~
\Phihat=\Phi^0+\Phi^1+\Phi^2\;\ldots +\Phi^n\ldots~~,
\en
where the fields $\Om^n$ and $\Phi^n$ are homogeneous polynomials in
$\theta$ of order $n$. 
If we multiply the differential equations by $\theta^{AB}$ and use the
identities
$\theta^{AB}\frac{\partial}{\partial
  \theta^{AB}}\Omhat^{n+1}=(n+1)\Omhat^{n+1}$ and
 $\theta^{AB}\frac{\partial}{\partial
   \theta^{AB}}\Phihat^{n+1}=(n+1)\Phihat^{n+1}$,
we obtain the recursive relations
\begin{flalign}
\Omhat^{n+1} &= \frac{i\,\theta^{AB}}{4(n+1)} \{\Omhat_A,  \ell_B \Omhat +\Rhat_B\}_\star^{n} \label{Omn+1} ~,\\
\Phihat^{n+1} &=\frac{i\,\theta^{AB}}{4(n+1)}  \{\Omhat_A,  \LL_B
\Phihat \}_\star^{n} \label{phin+1}~,
\\
\epsihat^{\;n+1} &=\frac{i\,\theta^{AB}}{4(n+1)}  \{\Omhat_A,  \ell_B
\epsihat \}_\star^{n} \label{epsin+1}~,
\end{flalign}
where for any field $P$ (also composite like  for ex.  $\{\Omhat_A,
\LL_B \Phihat \}_\star$),
$P^{n+1}$ denotes its component of
order $n+1$ in $\theta$.

\subsection{Differential equation and recursive relations for
 composite fields}\label{DERCF}

The $\theta$ expansion of  NC actions in terms of
classical fields can now in principle be carried out by expanding
the $\star$-products and the noncommutative fields 
by iterative use of the relations (\ref{Omn+1}) and
(\ref{phin+1}). The resulting formulae, already at second order in
$\theta$, become quite long, see for example the $\theta^2$ formulae for the Yang-Mills action in \cite{Moller} and the gravity action in
\cite{AC3}. They are not manifestly gauge invariant, and only repeated
integrations by parts allow to express the actions in terms of gauge
covariant quantities. 

A better strategy leads directly to an explicit gauge invariant $\theta$-expanded action: rather than expanding the $\star$-products and the
elementary fields present in the NC action, we first study the SW map for composite fields (i.e., $\star$-products of fields and their
derivatives), and find recursive relations in the noncommutativity
parameter $\theta$ as well as in the number of fields. A special case of
composite field is the action itself, and in the next section we give recursive and explicitly gauge invariant relations for the NC gravity action. The calculation of the expanded action at order $\theta^2$ then becomes straightforward.

\sk

In this section and in Section 4, for the  sake of notational simplicity,  we omit the hat that
denotes noncommutative fields, we also omit to write the $\star$ and $\wedge_\star$
products and simply write $\{~,~\}$, $[~,~]$  rather than
$\{~,~\}_\star$, $[~,~]_\star$. 
\sk

\noi {\bf{Lemma 1}}~ Let $P,Q$ be arbitrary exterior forms in the adjoint
representation. Then
\eq
\{\Om^{}_{[A},\LL^{}_{B]}P\}
Q+P\{\Om^{}_{[A},\LL^{}_{B]}Q\} + 2\ell^{}_{[A}P\ell^{}_{B]}Q
=\{\Om^{}_{[A},\LL^{}_{B]}(P Q)\} +2L^{}_{[A}P L^{}_{B]}Q\,.
\en
A
similar formula holds also if $Q$ transforms in the fundamental
representation (simply
 omit the second and third  curly brackets $\{~,~\}$).
Notice the algebraic character of this formula: it holds for any
associative product between the symbols $\Om_A, P, Q$.
The proof is by a straightforward calculation.
\sk
\sk

\noi {\bf{Lemma 2}}~ Let 
\begin{flalign}
\frac{\partial P\,}{\partial
\theta^{AB}}=\frac{i}{4}\{\Om_{[A},\LL_{B]}P\} +
P'_{[A\,B]}~,\\[.2em]
\frac{\partial Q\,}{\partial
\theta^{AB}}=\frac{i}{4}\{\Om_{[A},\LL_{B]}Q\} +
Q'_{[A\,B]}~,
\end{flalign}
where $P'^{}_{[A\,B]},  Q'^{}_{[A\,B]}$ are forms that
characterize  the deviation  from the differential
equation (\ref{diffeqphi}), \big(i.e., from 
$\frac{\partial P\,}{\partial
\theta^{AB}}=\frac{i}{4}\{\Om_{[A},\LL_{B]}P\}$\big). 
Then
\eq
\frac{\partial_{} (P Q)}{\partial
\theta^{AB}}=\frac{i}{4}\Big(\{\Om^{}_{[A},\LL^{}_{B]}(P  Q)\}
+2L^{}_{[A}P L^{}_{B]}Q
+P'^{}_{[A\,B]} Q+P
Q'^{}_{[A\,B]}\Big)~.\label{DECF}
\en
This result easily follows from the previous lemma and from  the
$\star$-product and $\wedge_\star$-product variation
$P\wedge_{\star_{\theta+\delta\theta}}Q=P\wedge_{\star_\theta} Q+
\frac{i}{2}\delta\theta^{AB}\ell_AP\wedge_\star \ell_BQ$.
\sk

\noi {\bf{Corollary 1}} (Recursive relation for products of composite fields)
\eq
(PQ)^{n+1}=\frac{i\,\theta^{AB}}{4(n+1)}\Big(\{\Om_A,\LL_B(P Q)\}
+2L_AP\, L_B Q
+P'_{[A\,B]} Q+P
Q'_{[A\,B]}\Big)^{n}~\label{RRCF}.
\en
{\sl Proof}~  By the same reasoning used to obtain the recursive solutions (\ref{Omn+1})-(\ref{epsin+1}).
\sk

The next lemma relates the covariant Lie derivatives to the
curvature tensor contracted along the commuting vector fields $X_A$
and $X_B$, defined by\footnote{Notice that if there exist local coordinates
where the commuting vector fields $X_A$ equal the  partial derivatives
$\partial_\mu$, then we have $i_\nu i_\mu R=R_{\mu\nu}$, in accordance
with $R=\frac{1}{2}R_{\mu\nu}dx^\mu \wedge dx^\nu$.}
\eq
R_{AB}\equiv i_Bi_AR~.\label{defRAB}
\en
The square of two exterior
covariant derivatives equals the curvature tensor $R$; similarly
\sk
\noi {\bf{Lemma 3} }  The commutator $[L_A,L_B]=L_AL_B-L_BL_A$ of two covariant Lie derivatives
is equal to minus the curvature tensor $R_{AB}$:
\eq
[L_A,L_B]=-R_{AB}\label{LLR}~.
\en 
For any exterior form $P$ that transforms in the adjoint we have 
\eq
\theta^{AB}L_AL_BP
=-\frac{1}{2}\theta^{AB}[R_{AB},P]\label{PLL}~.
\en
Another useful identity is
\eq
\theta^{AB}I\!\!L_A\Omega_B=\theta^{AB}R_{AB}\label{thetaLLOm}~.
\en
{\sl Proof} ~
It is easy to verify (\ref{LLR}) on fields $\psi$ that transform in the
fundamental representation $[L_A,L_B]\psi=-R_{AB}\psi$. The proof for fields in the
adjoint is equivalent to (\ref{PLL}) and is also straightforward.

In order to prove (\ref{thetaLLOm}) we calculate 
\begin{flalign}
R_{AB}&=i_B
i_AR=i_Bi_A(d\Om-\Om\wedge\Om)=i_B(\ell_A\Om-d\Om_A-\Om_A\Om+\Om\Om_A)\nn\\[.2em]
&=
\ell_A\Om_B-\ell_B\Om_A-\Om_A\Om_B +\Om_B\Om_A\nn\\
&=L_A\Om_B-\ell_B\Om_A
\end{flalign}
where we used the Cartan formula $\ell_A=i_Ad+di_A$ and, in the second
line, that $i_B\ell_A=\ell_Ai_B$ because the vector fields $X_A$ and
$X_B$ commute, and then again the  Cartan formula.
{}Formula (\ref{thetaLLOm}) now follows  immediately from the definition $I\!\!L_A=L_A+\ell_A$.
\sk
\sk
\noi {\bf{Corollary 2}}
\eq
\theta^{AB}\int Tr\Big( \{\Om_A,I\!\!L_B(PQ)\} +2L_AP\,L_B
Q\,\Big)=\,\theta^{AB}\int Tr \Big(\{R_{AB}, P\}Q\Big)~.
\en 
The proof easily follows integrating by parts and using the cyclic
property of $\int Tr\;$.
\sk

We now consider $P$ and $Q$ composite fields dependent on
the noncommutative connection $\Omega$ and on the matter fields $\Phi$.

Using the previous results, and  recalling that the contraction operators and the Lie derivative along
commuting vector fields commute, we find:
\begin{flalign}
&R^{n+1}
=\frac{i\,\theta^{AB}}{4(n+1)}\Big(\{\Om_A,\LL_B R\} -
[R_A,R_B]\Big)^{n}\label{Rreq}\\[.2em]
&(R\Phi\Phi)^{n+1}
=\frac{i\,\theta^{AB}}{4(n+1)}\Big(\{\Om_A,\LL_B(R\Phi\Phi)\} +2L_A
R\,L_B(\Phi\Phi)- [R_A,R_B]\Phi\Phi+2R\,L_A \Phi L_B \Phi\Big)^{n}\label{RVVreq}\\[.2em]
&((RR)_{AB}\Phi\Phi)^{n+1}
=\frac{i\,\theta^{CD}}{4(n+1)}\Big(\{\Om_C,\LL_D((RR)_{AB}\Phi\Phi)\} +2L_C
(RR)_{AB}\,L_D(\Phi\Phi) \label{RRVVreq}\\
&~~~~~~~~~~~~~~~~~~~~~~~~~~~~~~~~~~+2(L_C R\,L_D R)_{AB} \Phi\Phi-
\{[R_C,R_D],R\}_{AB}\Phi\Phi+2(RR)_{AB\,}L_C \Phi L_D \Phi\Big)^n\nn\\[.2em]
&(L_A\Phi)^{n+1}=\frac{i\,\theta^{CD}}{4(n+1)}\Big(\{\Om_C,\LL_D(L_A\Phi)\} +2\{R_{AC},L_D\Phi\}\Big)^n\label{LVreq}
\\[.2em]
&(R\,L_A\Phi L_B\Phi)^{n+1}
=\frac{i\,\theta^{CD}}{4(n+1)}\Big(\{\Om_C,\LL_D (R\,L_A\Phi L_B\Phi)\} 
+2L_CR\,L_D(L_A\Phi L_B\Phi) \label{RLVLVreq}\\
&~~~~~~~~~~~~~~~~~~~~~~~~~~~~~~~ -[R_C,R_D]\,L_A\Phi L_B\Phi
+2R[\{R_{AC},L_D\Phi\},L_B\Phi]
+2R\,L_C L_A\Phi_{\:} L_DL_B\Phi\nn
\Big)^{n}\\[.2em]
& (D\Phi)^{n+1}=\frac{i\,\theta^{CD}}{4(n+1)}\Big(\{\Om_C,\LL_D(D\Phi)\}
 - 2\{R_C,L_D \Phi \} \Big)^n \label{DVreq}
\end{flalign}
where we
have defined $[P,Q]=PQ-QP$, $\{P,Q\}=PQ+QP$ and $P_{AB}=i_Bi_AP$ for  exterior forms
$P$, $Q$. 
In deriving (\ref{LVreq}) we found it convenient to rewrite formula (\ref{Omn+1}) as 
$\Om^{n+1}=\frac{i\,\theta^{CD}}{4(n+1)}\big(\{\Om_C,\LL_D\Om\}
-\{\Om_C,d\Om_D\}\big)^n$, that implies
$\Om^{n+1}_A=\frac{i\,\theta^{CD}}{4(n+1)}\big(\{\Om_C,\LL_D\Om_A\}
-\{\Om_C,\ell_A\Om_D\}\big)^n$; then we used $\ell_A\LL_C\Phi=-[\ell_A\Om_C,\Phi]+\LL_C\ell_A\Phi$.

\section{Recursive relations for the gravity action}
A particular composite field is also  the Lagrangian itself, and we calculate here
the SW recursive relation for the noncommutative vierbein gravity
action (\ref{NCaction}) . In this case we set $\Phi=V$.
Then  Corollary 2, formula (\ref{RVVreq}) and the identity
$(RR)_{AB}=\{R_{AB},R\}-[R_A,R_B]$ 
lead to the recursive relation
\eq
 S^{n+2} = \frac{-\theta^{AB}}{4(n+2)} \int Tr \left(\ga_5 \big( (RR)_{AB}VV+2R\,L_A V L_B V\big)_{}\right)^{n+1} ~\label{SRrel}
\en
where, as in Section \ref{DERCF},  we omit writing wedge and star products, and hats on
noncommutative fields.

 We now use (\ref{RRVVreq}) and (\ref{RLVLVreq}) and further obtain
 \begin{flalign}\label{Sn+2}
S^{n+2} &= \frac{ i\,\theta^{AB} \theta^{CD} }{4(n+2)4(n+1)}\!\int \!Tr\Big(\ga_5 \big( 
\{[R_C,R_D],R\}_{AB} - \{R_{CD}, (RR)_{AB}\} - 2(L_C R \,L_D R)_{AB}\big)VV\nonumber \\[.2em]
& ~~~~~~~~~~~~~~ - 4\gamma_5\big( (RR)_{AB} L_C V\,L_D V  -R\, [ \{ R_{AC} , L_B V \}
, L_D V] +R (L_A L_C V) (L_B L_D V) \big) \Big)^n~.
\end{flalign}
Other equivalent expressions for $S^{n+2}$ can  be obtained by using
the identities:
 \begin{flalign}
&(RR)_{CD}=\{R_{CD},R\}-[R_C,R_D]~,\nn\\[.4em]
&\{\{R_{AB},R\},R\}_{CD} - 2\{R_{CD},
  (RR)_{AB}\}=2(R\,R_{AB}R)_{CD}-\{(RR)_{AB},R_{CD}\}~.\nn
\end{flalign} 
\sk

\noi{\bf Note}~
The general recursive relation for composite fields (\ref{RRCF})
follows from the corresponding differential equation
(\ref{DECF}). Similarly the recursive relations
(\ref{Rreq})-(\ref{SRrel})  are also implied by corresponding
differential equations. In particular the action 
satisfies:
\eq
 \frac{\partial }{\partial \theta^{AB}} \,S = -\frac{1}{4} \int Tr \left(\ga_5 \big( (RR)_{AB}VV+2R\,L_A V L_B V\big)_{}\right) ~.
\en

\sect{Extended gravity action at $\theta^2$}
If we set $n=0$ in (\ref{Sn+2}) we obtain the expression for $S^2$
where only
commutative fields and usual undeformed products appear. Since in \cite{AC3} we have
shown that the expansion of the noncommutative gravity action in  the
commutative vierbein and spin connection fields contains only even
powers of $\theta$, then $S^2$ is the first nontrivial term in
this expansion:
\eq
  \int Tr (i \ga_5 \Rhat \westar \Vhat \westar \Vhat  )=
  \int Tr (i \ga_5 R \we V \we V  )+S^2+O(\theta^4)~.
\en
Finally, inserting the classical gamma expansions
(\ref{classicalfields}) of $V$ and $R$ and taking the trace  brings
the second order correction to the explicit form (repeated Lorentz
indices are summed with the Minkowski metric $\eta_{ab}$; the wedge
product between forms is omitted):
 \begin{flalign}
&   S^2 = {1 \over 32} \theta^{AB} \theta^{CD} \int  - \unmezzo \big((
R^{ab}_C R^{cd}_D R^{ef})^{}_{AB} - \unmezzo R^{ef}_{CD} (R^{ab} R^{cd} )^{}_{AB} \big)
V^g V^h (\epsi_{abcd} \de^{gh}_{ef} + \epsi_{efgh} \de^{cd}_{ab}
)\nonumber \\[.3em]
& ~~~~~~~~~~~~~~~~~~~~~~~~~ + \big( 2 (L_C R^{\,ea} L_D R^{\,eb})^{}_{AB} V^c V^d - (R^{ab} R^{cd} )^{}_{AB}
L_C V^{\,e} L_D V^{\,e}  \nonumber \\[.3em]
& ~~~~~~~~~~~~~~~~~~~~~~~~~~~~~~ - 8 R^{df} R^{ab}_{AC} L_B V^{\,c}
L_D V^{\;\!f} -  4 R^{ab} (L_A
  L_C V^{\,c})(L_B L_D V^{\,d})\big) \epsi_{abcd} \,~.
 \end{flalign}
\sk
We conclude by observing that it is also possible to expand the noncommutative 
cosmological term $\frac{-i}{4!}\int Tr(\gamma_5\Lambda_0\Vhat\westar
\Vhat\westar\Vhat\westar\Vhat)$ in power series of $\theta$. 
Following the study in \cite{AC3} on charge
conjugation symmetry and parity under $\theta\to -\theta$,  we again have that only even
powers of $\theta$ contribute. We find:
\eq
\frac{-i}{4!}\int Tr(\gamma_5\Lambda_0\Vhat\westar
\Vhat\westar\Vhat\westar\Vhat)=
\frac{-i}{4!}\int Tr(\gamma_5\Lambda_0V\we
V\we V \we V) +S^2_{\Lambda_0}+O(\theta^4)~
\en
where 
\begin{flalign}
S^2_{\Lambda_0}=\:&\frac{i\Lambda_0 \,\theta^{AB}\theta^{CD}}{4!\; 4}\int
Tr\Big(\gamma_5\big(\,(\frac{1}{2}R_{AB}R_{CD}-R_{AC}R_{BD})VVVV +\frac{1}{2}R_{AB}L_C(VV)L_D(VV)\nn\\[.2em]
&~~~~~~~~~~~~~~~~~~~~~~~~~~\,~~~~+R_{AB}\{VV,L_CVL_DV\}+L_AVL_BVL_CVL_DV\nn\\[.2em]
&~~~~~~~~~~~~~~~~~~~~~~~~~~\,~~~~-VV[\{R_{AC},L_BV\},L_DV] 
+VV(L_AL_CV)(L_BL_DV)_{}\big)\Big)~.
\end{flalign}
This expression agrees with the result in \cite{AC5}, Section 6, where
the expansion of the NC cosmological term is obtained as the
$\varphi\to 1$ limit of the $\theta$-expansion of a term $\int
Tr(i\gamma_5\varphi\varphi VVVV)$ describing scalars $\varphi$ in a
curved background.

\sk
\sk
\noi {\bf Acknowledgements}\\
M. D. thanks the INFN sezione di Torino, gruppo
collegato di Alessandria, for  hospitality during her visit. The work of M.D. is supported by
Project No.171031 of the Serbian Ministry of Education and Science.
\appendix
\sect{Cartan formulae}
The usual Cartan calculus formulae simplify if we consider commuting
vector fields $X_A$, and read 
 \begin{flalign}
&~~~~~~~~~~~~~~~~\ell_A=i_Ad+di_A ~,& &~~L_A=i_AD+D i_A&\\
&~~~~~~~~~~~~~~~~[\ell_A,\ell_B]=0~,& &~~[L_A,L_B]=i_Ai_B R&\\
&~~~~~~~~~~~~~~~~[\ell_A,i_B]=0~,& &~~[L_A,i_B]=0&\\
&~~~~~~~~~~~~~~~~i_Ai_B+i_Bi_A=0~,~~~~~~d\circ d=0~,\!\!\!\!\!\!\!\!\!\!\!\!\!\!\!\!\!\!\!\!\!\!\!\!\!\!\!\!\!\!\!\!\!\!\!\!\!\!\!\!\!\!  & &~~D\circ D= R&
\end{flalign}

\sect{Gamma matrices in $D=4$}

We summarize in this Appendix our gamma matrix conventions in $D=4$.
\begin{flalign}
&  \eta_{ab} =(1,-1,-1,-1),~~~\{\ga_a,\ga_b\}=2 \eta_{ab},~~~[\ga_a,\ga_b]=2 \ga_{ab}, \\
&  \ga_5 \equiv i \ga_0\ga_1\ga_2\ga_3,~~~\ga_5 \ga_5 = 1,~~~\epsi_{0123} = - \epsi^{0123}=1, \\
&  \ga_a^\dagger = \ga_0 \ga_a \ga_0, ~~~\ga_5^\dagger = \ga_5 \\
&  \ga_a^T = - C \ga_a C^{-1},~~~\ga_5^T = C \ga_5 C^{-1}, ~~~C^2 =-1,~~~C^\dagger=C^T =-C
\end{flalign}

\subsection{Useful identities}

\begin{flalign}
 & \ga_a\ga_b= \ga_{ab}+\eta_{ab}\\
 &  \ga_{ab} \ga_5 = {i \over 2} \epsilon_{abcd} \ga^{cd}\\
 & \ga_{ab} \ga_c=\eta_{bc} \ga_a - \eta_{ac} \ga_b -i \epsi_{abcd}\ga_5 \ga^d\\
 & \ga_c \ga_{ab} = \eta_{ac} \ga_b - \eta_{bc} \ga_a -i \epsi_{abcd}\ga_5 \ga^d\\
 & \ga_a\ga_b\ga_c= \eta_{ab}\ga_c + \eta_{bc} \ga_a - \eta_{ac} \ga_b -i \epsi_{abcd}\ga_5 \ga^d\\
 & \ga^{ab} \ga_{cd} = -i \epsi^{ab}_{~~cd}\ga_5 - 4 \de^{[a}_{[c} \ga^{b]}_{~~d]} - 2 \de^{ab}_{cd}\\
&  Tr(\ga_a \ga^{bc} \ga_d)= 8~ \de^{bc}_{ad} \\
&  Tr(\ga_5 \ga_a \ga_{bc} \ga_d) = -4 i\,\epsi_{abcd} 
 \end{flalign}
\sk
\noi where
$\delta^{ab}_{cd} \equiv
\frac{1}{2}(\delta^a_c\delta^b_d-\delta^b_c\delta^a_d)$, 
and indices antisymmetrization in square brackets has total weight $1$.

\end{document}